%% file: NJP_LiCs.tex
\documentclass[12pt]{iopart}
\usepackage{graphics}
\usepackage{rotating}
\usepackage{epsfig}
\usepackage{color}
\newcommand{\ket}[1]{\vert{#1}\rangle}

\begin{document}

\title{Ultracold molecules: vehicles to scalable quantum information
processing}

\author{Kathy-Anne Brickman Soderberg, Nathan Gemelke, and \\ Cheng Chin}

\address{James Franck Institute and Physics Department, University of Chicago,\\
  Chicago, IL 60637}
\ead{cchin@uchicago.edu}
\begin{abstract}
We describe a novel scheme to implement scalable quantum information processing using Li-Cs molecular state to entangle $^{6}$Li and $^{133}$Cs ultracold atoms held in independent optical lattices.  The $^{6}$Li atoms will act as quantum bits to store information, and $^{133}$Cs atoms will serve as messenger bits that aid in quantum gate operations and mediate entanglement between distant qubit atoms. Each atomic species is held in a separate optical lattice and the atoms can be overlapped by translating the lattices with respect to each other.   When the messenger and qubit atoms are overlapped, targeted single spin operations and entangling operations can be performed by coupling the atomic states to a molecular state with radio-frequency pulses.  By controlling the frequency and duration of the radio-frequency pulses, entanglement can either be created or swapped between a qubit messenger pair.  We estimate operation fidelities for entangling two distant qubits and discuss scalability of this scheme and constraints on the optical lattice lasers.
\end{abstract}

\maketitle

\input{intro_KA.tex}

\input{scheme_KA.tex}


\input{entanglement.tex}

\input{lattice.tex}

\input{conclusion.tex}

\section{Acknowledgements}
The authors wish to thank Skyler Degenkolb and Andreas Klinger for
production of optical elements and lattice construction as shown in
Figs.~\ref{lattice_perf} and \ref{lattice_optics}, Arjun Sharma,
Scott Waitukaitis, Kara Lamb, and Peter Scherpelz for construction
of dual-species apparatus.  The authors acknowledge support from the
NSF-MRSEC program under No. DMR-0820054 and ARO Grant No.
W911NF0710576 from the DARPA OLE Program and Packard foundation.
N.G. acknowledges support from the Grainger Foundation.  K.-A.B.S.
acknowledges support from the Kadanoff-Rice MRSEC Fellowship.

\end{document}

%% file: intro_KA.tex
\section{Introduction}
\label{Intro}
The production of scalable, controlled quantum entanglement between many particles would represent a revolutionary breakthrough for information processing.  Shortly after Shor's famous algorithm \cite{shor1997} proved in principle that a quantum computer could factor large numbers exponentially faster than any current classical algorithm, there was an exponential growth in the number of proposals for how to implement the essential elements of quantum computation.  Since then, many systems have made great strides toward realizing such a computer \cite{monroe2002,blatt2008, vandersypen2005, you2005,hanson2007, raimond2001, pryde2008, kok2007,jessen2004, prevedel2007}, however, truly scalable information processing remains an elusive goal.  This is due in part to the stringent requirements on long coherence times, the technical difficulties in implementing high fidelity entangling operations, and the challenge to store and control interactions between many quantum bits (qubits). While neutral atoms provide a natural advantage in coupling weakly to their environment and to other atoms at long distance,  atomic interactions at short-range, well described by contact interactions, can be strong, coherent, and their effect can be controlled by overlapping the atomic wavefunctions.  In particular, the strength of this contact interaction is highly sensitive to underlying molecular structure, and can be precisely manipulated by introducing direct coupling mechanisms between free atoms and molecules.

A system using both ultracold molecules and atoms held in an optical lattice may be a promising system to realize a scalable quantum computer due to the high degree of control available in these systems \cite{daley2008,rabl2006}.  Many recent theoretical proposals present schemes to implement entangling operations with neutral atoms in optical lattices \cite{jaksch1999,brennen1999,jaksch2000} and several experimental groups have demonstrated key steps towards the goal of quantum information processing \cite{greiner2002,mandel2003,peil2003,bloch2008, scheunemann2000,nelson2007, johnson2008, urban2008,anderlini2007,trotzky2008}.  Atoms trapped in optical lattices in particular lend themselves to scalability because thousands of atoms can be isolated in a regular array of micron-sized volumes.  Atoms localized in the ground state of each site in the tight-binding regime provide an excellent environment to store quantum information with long coherence times $T_{coh}>1$s \cite{boyd2006} and can be spatially transported by controlling the optical phases of the lattice beams \cite{mandel2003,miroshnychenko2006}.  The proposal presented here is a novel approach to use two atomic species, each manipulated by a separate optical lattice potential.  Highlighted is the fabrication of lattice structure independent of optical wavelength, use of molecular states to induce entanglement between atoms, and introduction of single site addressability without the need for spatially resolved manipulations.

A key aspect of this approach is the introduction of auxiliary messenger atoms used both to probe and to manipulate quantum states and entanglement in an array of qubit atoms.  By utilizing two separate species of atom for these two roles, and carrying information in their internal states, it becomes technically feasible to manipulate spatial overlap of atoms and thereby their interactions, without disrupting the sensitive quantum coherences.  We propose to use fermionic $^{6}$Li atoms as qubits, prepared in the lattice with ideally one atom per site.  Bosonic $^{133}$Cs will act as messenger atoms to aid in the gate operations and mediate entanglement among the qubits, and will be less densely populated, on order of one atom per 100 sites of a separate lattice potential of identical structure to the first.  By shifting the relative alignment of the lattices through optical phases, each $^{133}$Cs atom can in principle be transported to any distant $^{6}$Li atom; similar transport schemes can be found in Ref.~\cite{you2000,cirac2000,chin2001}.  Since there may be many $^{133}$Cs atoms, multiple copies of the same computation can proceed in parallel. 

%% file: scheme_KA.tex
\section{Scalable quantum information processing with atoms and molecules in optical lattices}
\label{scheme}

The necessary requirements to implement a scalable quantum computer include the ability to initialize the qubit register, fabricate a universal gate set, to have long decoherence times, and to read out the information \cite{divincenzo2000}.  This section will outline our proposal to meet these requirements.

\begin{figure}[htbp]
\begin{center}
\includegraphics [width=0.75\columnwidth,keepaspectratio]{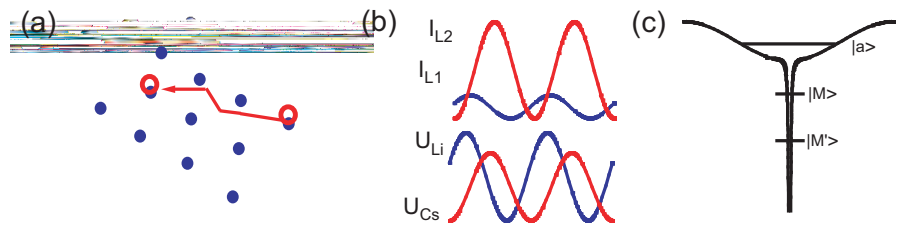}
\caption{Scheme for scalable quantum information processing in optical lattices. (a) qubit atoms (blue dots) form a band insulator in the optical lattices with unity occupancy. Entanglement of two distant qubits can be mediated by the messenger atom (open red circle), which is controlled by a second set of optical lattices (not shown, see text).  (b) Top shows offset intensty profiles for 681~nm light (blue) and 1064~nm (red) light, here $I_{\rm Li}=0.24I_{\rm Cs}$.  Bottom shows resulting potentials for $^6$Li (blue) and $^{133}$Cs (red).  (c) Potential energy in the center-of-mass coordinate, including the $^6$Li-$^{133}$Cs interatomic interaction.  $\ket{a}$ represents an atomic state, and $\ket{M}$ and $\ket{M'}$ represent molecular states.}
\label{qcscheme}
\end{center}
\end{figure}

In recent years researches working on neutral atom optical lattice experiments have made great progress obtaining complete quantum control over atoms in a lattice \cite{bloch2008}. An optical lattice is the intensity pattern of several interfering laser beams; the resulting periodic pattern can be shaped by varying the intensity, propagation directions, optical phases, and polarization of the laser beams. For an effectively two-level atom with far-detuned laser beams, the potential $V(\vec{x})$ is given by $V(\vec{x})= \frac{\hbar\Gamma}8 \frac{I(\vec{x})/I_{sat}}{\Delta/\Gamma}$, where $\Gamma$ is the natural linewidth of the atomic transition, $I_{sat}$ is the saturation intensity, and $\Delta=\omega-\omega_0\gg\Gamma$ is the laser detuning from resonance at $\omega_0$.  $I(\vec{x})$ is the intensity of the optical lattice.

Our scheme to implement a scalable quantum information processor is sketched in Fig.~\ref{qcscheme}. Two sets of three-dimensional lattices will confine each atomic species independently.  One lattice at a wavelength $\lambda_1$=681~nm will primarily affect confinement of $^6$Li qubit atoms, whose filling ratio will be near unity.  A second, less densely populated lattice, at $\lambda_2$=1064~nm will hold $^{133}$Cs messenger atoms to serve as auxiliary quantum bits which allow single site addressing of the qubits, carry entanglement between qubit atoms, and enable readout operations.  The lattices will be fabricated by tuning intersection angles to have identical lattice potential spacings for each species, achieved by using diffractive optical techniques described in Sec.~\ref{lattice}.  Additionally, one lattice will be physically translatable, to allow controlled contact between qubit and messenger atoms.

\begin{figure}[htpb]
\begin{center}
\includegraphics [width=0.6\columnwidth,keepaspectratio]{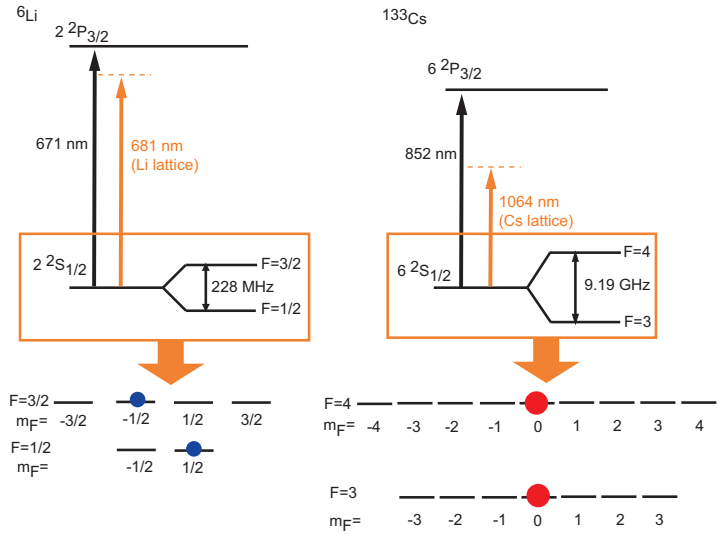}
\caption{Energy level diagrams of $^{6}$Li and $^{133}$Cs showing the relevant transitions and qubit levels.  The qubit levels in $^6$Li are $\ket{F=3/2,m_{F}=-1/2} \equiv \ket{1}_{\rm Li}$ and $\ket{F=1/2,m_{F}=1/2} \equiv \ket{0}_{\rm Li}$, they are denoted by blue circles.  In $^{133}$Cs, the qubit levels are the ground state hyperfine clock states, $6 ^{2}S_{1/2} \ket{F=4,m_{F}=0} \equiv \ket{1}_{\rm Cs}$ and $6 ^{2}S_{1/2} \ket{F=3,m_{F}=0} \equiv \ket{0}_{\rm Cs}$ and are denoted by the red circles.  The orange arrows denote the lattice laser wavelengths of $\lambda_1$ = 681~nm and $\lambda_2$ = 1064~nm.}
\label{elevel}
\end{center}
\end{figure}

The choice of fermionic $^6$Li permits high fidelity intialization of the lattice with one atom per site, achieved by increasing the lattice depth to induce a band insulator state in the atoms \cite{kohl2005}. The energy levels for $^6$Li and $^{133}$Cs are shown in Fig.~\ref{elevel}.  We propose to use the ground state hyperfine levels $\ket{F=3/2, m_{F}=-1/2}\equiv \ket{1}_{\rm Li}$ and $\ket{F=1/2, m_{F}=1/2}\equiv \ket{0}_{\rm Li}$ of $^6$Li; here $F$ is the total angular momentum, and $m_F$ its projection along the quantization axis.  These states are chosen because they have the same magnetic moment, to minimize decoherence due to external fields.

In $^{133}$Cs, the ground state hyperfine `clock' states  $\ket{F=4, m_{F}=0}\equiv \ket{1}_{\rm Cs}$ and  $\ket{F=3, m_{F}=0}\equiv \ket{0}_{\rm Cs}$ will be used, see Fig.~\ref{elevel}.  These states are insensitive to magnetic field to first order.

Independent control of the qubit and messenger atoms is essential in this setup and can be realized by a careful choice of the lattice laser detunings and intensities. The choice of $^6$Li and $^{133}$Cs is favorable in this sense due to their very different dominant atomic transition lines at $\lambda$=671~nm and 852~nm, respectively. This opens up the possibility to independently confine $^6$Li and $^{133}$Cs atoms with two sets of moderately detuned optical lattices $L_1$ and $L_2$.

\begin{figure}[htpb]
\begin{center}
\includegraphics [width=0.6\columnwidth,keepaspectratio]{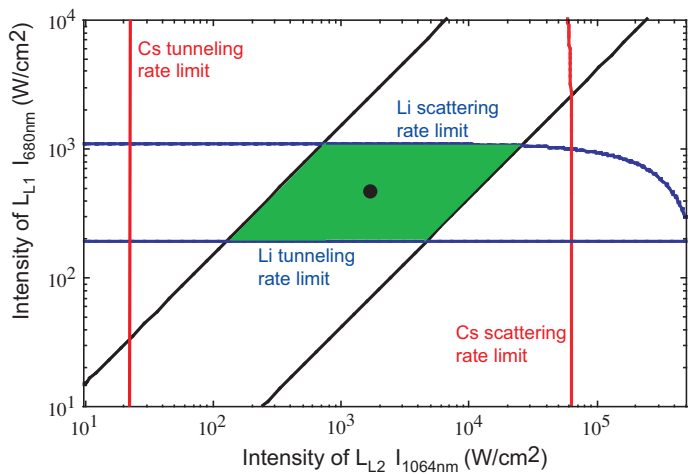}
\caption{Plot of limits on the intensities of the 681~nm optical lattice $L_1$ vs. the 1064~nm optical lattice $L_2$.  The diaganol black lines show the bounds imposed by requiring independent control of $L_1$ over $^6$Li and $L_2$ over $^{133}$Cs.  Also shown are the tunneling rate limit and off-resonant scattering rate limit for both $^6$Li (blue lines) and $^{133}$Cs (red lines) for a decoherence rate of 2/s.  The green shaded box shows the available parameter space satisfying all of the above conditions.  The black dot corresponds to conditions assumed for calculations in the text.}
\label{intensityratio}
\end{center}
\end{figure}

Figure \ref{intensityratio} illustrates the constraints imposed on lattice intensities to maintain independent control of $^6$Li and $^{133}$Cs, and bit lifetime of 500~ms due to off-resonant scattering and tunneling. The lattice spacing was chosen to be 1.5~$\mu$m.  By evaluating the maximum force each lattice can exert on each of the two atomic species, we show that the condition for independent control of the atoms by the associated lattices can be expressed in terms of the lattice laser intensities, which should satisfy $0.04<I_{1}/I_{2}<1.45$. With the optimal choice of $I_{1}/I_{2}=0.24$, the dipole force from the farther detuned lattice does not exceed $\alpha=16\%$ of the force from the nearer detuned lattice.  In addition, this plot shows the limits for two primary decoherence sources, off-resonant scattering of lattice light and atomic tunneling.  In both cases, tighter constraints are set by the $^6$Li atoms due to their lighter mass and to the smaller detuning of $L_1$ to the $^6$Li transition.

The capability to independently control the two atomic species
allows us to have single site addressability of the qubit atoms.
This is accomplished by shifting the optical phases of the messenger
lattice, allowing the $^{133}$Cs messenger atom to be translated to
any $^6$Li qubit atom.  This is a necessary step for many operations
in this proposal, including detection and creation of a universal
gate set, see Sec.~\ref{entanglement}.

There are several possibilities for reading out the quantum
information from the qubits.  One approach is to use the auxiliary
messenger $^{133}$Cs atoms to determine the state of the $^6$Li
atoms, allowing information to be obtained without disturbing the
qubit lattice.  A second approach is to readout the state of the
qubit lattice directly.  This could be done by using high
numerical-aperture, state-selective imaging of the qubit lattice
directly, for example, see Ref.~\cite{nelson2007}.

\begin{figure}
\begin{center}
\includegraphics [width=0.6\columnwidth,keepaspectratio]{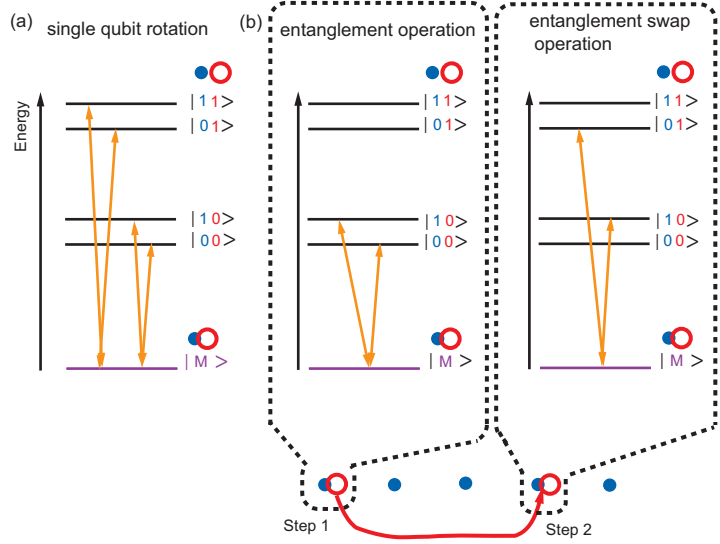}
\caption{Implementation of the necessary gates via coupling to a molecular state $\ket{M}$.  Not only does the molecular state allow entangling operations to be carried out between the atoms, but it also allows single qubit addressing.  Part (a) shows how to execute a targeted single qubit rotation.  When the atoms are overlapped, radio-frequency pulses allow the qubit atom to be rotated into a superposition state.  Part (b) shows the sequence to entangle two distant qubits.  After entangling messenger and qubit (Step 1), the messenger can be transported to a second qubit and subsequently entanglement can be exchanged (Step 2).}
\label{gates}
\end{center}
\end{figure}

The most fundamental requirement for realizing a quantum information
processor is the need for a universal gate set.  This includes both
single qubit rotation gates and multi-qubit entangling operations.
One possible way to achieve a universal gate set in this system is
shown in Fig.~\ref{gates}.  In both cases the logic states of the
atoms can be coupled to a $^6$Li-$^{133}$Cs molecular state through
the use of radio-frequency (rf) fields.

To perform targeted qubit rotations, where only a single qubit is
rotated and the neighboring qubits remain unaffected, we will
translate the messenger atoms by shifting the $L_2$ lattice in order
to overlap the messenger with the target qubit.  When the two atoms
are overlapped, they can be coupled to a molecular state using rf
transitions and, depending on the frequency and duration of the rf
pulses, we can perform any arbitrary Bloch rotation, see
Fig.~\ref{gates}(a).  Global rotations of all qubits can be realized
with microwave pulses.

Figure~\ref{gates}(b) shows a possible protocol to entangle two
distant qubits.  In the first step, the messenger, prepared in a
superpostion state, is brought to the first qubit and entangled by
rf pulses as shown in Step 1 in Fig.~\ref{gates}(b).  Next, the
messenger atom is translated to the second qubit and the quantum
entanglement is swapped between the messenger and qubit, as shown in
Step 2 in Fig.~\ref{gates}. This leaves the qubit atoms entangled
with each other and the messenger atom disentangled from the qubits.
The overall evolution of the quantum states for entangling two
distant qubits, Li$^a$ and Li$^b$, via the a messemger Cs atom is
given as

\begin{eqnarray}
&|{\rm Cs}:0+1\rangle\bigotimes|{\rm Li^a:0}\rangle\bigotimes|{\rm
Li^b:0}\rangle \,\,\,\,\,\overrightarrow{{\rm Step
1}}\,\,\,\,\,(-|01\rangle+|10\rangle)\bigotimes|0\rangle
\\
&\overrightarrow{{\rm Step2}}\,\,\,\,\,
-|010\rangle-|001\rangle=-|0\rangle\bigotimes(|10\rangle+|01\rangle).
\label{entanglement}
\end{eqnarray}


%% file: entanglement.tex
\section{Entanglement via quantum states of ultracold $^6$Li-$^{133}$Cs molecules}
\label{entanglement}

Molecular states are excellent candidates to induce entanglement of atoms because the molecular potential in general depends on atomic spin. In this section, we evaluate the times and fidelities to induce single spin rotations
and entanglement of qubits via radiative transitions to molecular
states as shown in Fig.~\ref{gates}(b).

There are two distinct ways that two free atoms can couple to the
molecular states. First, coupling to deeply-bound $^6$Li-$^{133}$Cs molecules can be
induced by direct radiative (electric dipole $D1$) transitions. Alternatively,
coupling to weakly-bound molecules near the continuum can be realized using rf or microwave
transitions (magnetic dipole $M1$). In the following paragraphs, we will focus on the magnetic dipole transitions.

In the small binding energy limit, the Rabi
frequency for magnetic dipole transitions between a free-atom state and a molecular state can be
estimated as $\Omega=\Omega_0 C$, where $\Omega_0$ is the typical Rabi frequency for the transition for free-atoms, and the Franck-Condon factor $C=\int{\psi_a(r)\psi_m(r)dr}$ is given by the wavefunction overlap
of the atomic state $\psi_a(r)$ and the molecular state $\psi_m(r)$.

To address a $^6$Li atom in the lattice, the $L_2$ is translated to bring a $^{133}$Cs atom into wavefunction overlap with $^6$Li. The Hamiltonian of the two atoms in a lattice site is characterized by one $^6$Li
with mass $m_1$, momentum $p_1$, position $r_1$, trap frequency
$\omega_1$ and one $^{133}$Cs atom with $m_2$, $p_2$, $r_2$ and $\omega_2$.
The Hamiltonian in the lab coordinate and in the center-of-mass frame are given by

\begin{eqnarray}
H&=&\frac{p_1^2}{2m_1}+\frac{p_2^2}{2m_2}+\frac{m_1\omega_1^2r_1^2}2+\frac{m_2\omega_2^2r_2^2}2+V(|r_1-r_2|)
\\ &=&\frac{P^2}{2M}+\frac{p^2}{2\mu}+\frac{M\omega_c^2R^2}2+\frac{\mu \omega_r^2r^2}2+V(r),
\label{eq:mol_LZ}
\end{eqnarray}

\noindent respectively, where $M=m_1+m_2$ is the total mass,
$\mu=m_1m_2/(m_1+m_2)$ is the reduced mass, $P=p_1+p_2$ is the total
momentum, $p=\mu(p_1/m_1-p_2/m_2)$ is the relative momentum,
$r=|r_1-r_2|$ is the atomic separation,
$\omega_c^2=(m_1\omega_1^2+m_2\omega_2^2)/M$ is the center-of-mass
trap frequency and $\omega_r=\omega_1\omega_2/\omega_c$ is the
relative motion trap frequency.

In the relative coordinate, the ground state wavefunction of a
weakly-interacting atom pair is given by
$\psi_a(r)=(r_0^2\pi)^{-3/4}\exp(-r^2/2r_0^2)$, and
$r_0=(\hbar/\mu\omega_r)^{1/2}$ is the oscillator length. The wavefunction of a weakly-bound molecular state is $\psi_m(r)=(2\pi ar)^{-1/2}\exp(-r/a)$,
where we assume the $^6$Li-$^{133}$Cs scattering length $a$ is larger than the
interaction range, but small compared to the trap oscillator length.
We can then evaluate the integral and obtain the desired
atom-molecule Rabi freuqnecy as
$\Omega=2\pi^{-1/4}(a/r_0)^{3/2}\Omega_0$. This result suggests that
the coupling rate can be enhanced for large scattering lengths. Given the trap parameters
described in Sec.~\ref{scheme}, and assuming a typical scattering length of
$a=200a_B$ (with $a_B$ Bohr radius), and atomic Rabi frequency of $\Omega_0=2\pi\times$10~kHz, we find $\omega_r=2\pi\times$160~kHz, $r_0=210$~nm, and
$\Omega=2\pi\times$200~Hz.  As discussed above, both single spin rotations and $^6$Li-$^{133}$Cs entanglement operations require two $\pi$ pulses on the atom-molecule transition, and consequently require a total operation time of
$\tau=\pi/\Omega=2.5$~ms.

The fidelity of the above operations can be estimated from the
uncertainty of the coupling strength and off-resonant population transfer to other states. One major source of the
coupling strength variation comes from the imperfect overlap of the
$^{133}$Cs and $^6$Li ground state wavefunctions, particularly when the lattice
site positions are not perfectly controlled. In Sec.~\ref{lattice}, we show
that the relative lattice positions can likely be controlled to about
$\delta$=10~nm in the near future. By evaluating the wavefunction
overlap with such an offset, we derive the resulting fidelity $\mathcal{F}$, defined as the square of the overlap between target and actual output states, to be
$\mathcal{F}=\exp(-\delta^2/r_0^2)=99.5\%$ per operation. For both spin rotations
and $^6$Li-$^{133}$Cs entanglement, which require two atom-molecule transitions
(see Sec.~II), we expect the overall fidelity to be $99\%$. The
dominant off-resonant population transfer will occur when the atoms are excited
to unintended molecular states, or the molecule is converted into
atoms in other vibrational states. From a two-level model, the off-resonant
population transfer is given as $\delta p=(1+\Delta^2/4\Omega^2)^{-1/2}$
per $\pi$-pulse, where the smallest detuning is determined by the
lattice vibrational energy of $\Delta=\omega_r=2\pi \times$160~kHz, which
suggests $\delta p=0.2\%$. The loss in fidelity from off-resonant
excitation is thus likely smaller than that from the lattice misalignment.

Entanglement of two qubits requires time not only to
entangle $^6$Li and $^{133}$Cs at a lattice site, but also to transport the
cesium atoms to a third, distant, $^6$Li atom and to transfer entanglement, see Sec.~\ref{scheme}. The latter process requires two atom-molecule $\pi$-pulses,
which take 2.5~ms with a fidelity of 99$\%$, similar to
the $^6$Li-$^{133}$Cs entanglement gate. In the following, we
estimate the time required to adiabatically transport a cesium atom
over $N$ lattice sites with fidelity $\mathcal{F}=99\%$.

We will adopt an adiabatic transportation process to keep messenger atoms in the vibrational
ground state while moving them. The leading order
loss of quantum information comes from the population transfer from
the ground state $|0\rangle$ to an excited state
$|i\rangle$. Based on an adiabatic approximation\cite{schiff}, we estimate

\begin{equation}
p_i=\frac1{\hbar^2\omega_{i}^4} |\langle i|\frac{\partial
H(t)}{\partial t}|0\rangle e^{i\omega_{i}t}|^2\rho(E)\omega_i t ,
\label{eq:adiabaticity}
\end{equation}

\noindent where $\rho(E)=1/\hbar\omega$ is the density of states in
the direction of motion,and the time-dependent $H(t)=U^* \sin^2k(x-vt)$ in
the frame moving with the cesium atoms is mostly due to the $L_1$ lattice potential, $U^*$ is the maximum $L_1$ lattice depth experienced
by the cesium atoms, and $v$ is the velocity of $L_1$ relative to $L_2$. When a cesium atom
moves over $N$ lattice sites, we can express the population transfer to
the lowest excited state $i=1$ as

\begin{equation}
p_1=N\frac{\pi}2 \nu
\frac{U^*}{\hbar^2\omega^2}k^2x_0^2 e^{-k^2x_0^2},
\label{eq:adiabaticity2}
\end{equation}

\noindent where we have introduced the reduced velocity $\nu=v\pi
\omega/k$. Using the parameters described
in Sec.~\ref{scheme}, $U^*=\alpha \hbar \times 760$~kHz with the
cross-talk suppression factor $\alpha=0.16$, $k=\pi/d$ and $d=1.5$~$\mu$m, we
derive, for a fidelity $99\%$, $\nu=0.03/N$, which corresponds to an
upper bound of the cesium atomic velocity of $v\approx(4/N)\mu$m/ms when it is required to move over $N$ sites.

To summarize, entanglement of two distant $^6$Li atoms separated by $N$
lattice sites involves four atom-molecule transitions and one
lattice transport. We conclude that the total time needed to perform
the whole sequence is $\tau_e=(5+0.4N^2)$ms with an overall fidelity
of 97$\%$. Since the total number of accessible qubits within a
travel distance equivalent of $N$ lattice sites is
$N_q=(4\pi/\sqrt{3}) N^2$, the time needed to realize a single pairwise entanglement gate in a system of $N_q$ qubits is $\tau_e=(5+N_q/20)$~ms on average. The
weak dependence on $N_q$ confirms that our entanglement scheme is
indeed scalable to many qubits.

%% file: lattice.tex
\section{Controlled overlap bichromatic optical lattices}
\label{lattice}

\begin{figure}[htpb]
\begin{center}
\includegraphics [width=0.95\columnwidth,keepaspectratio]{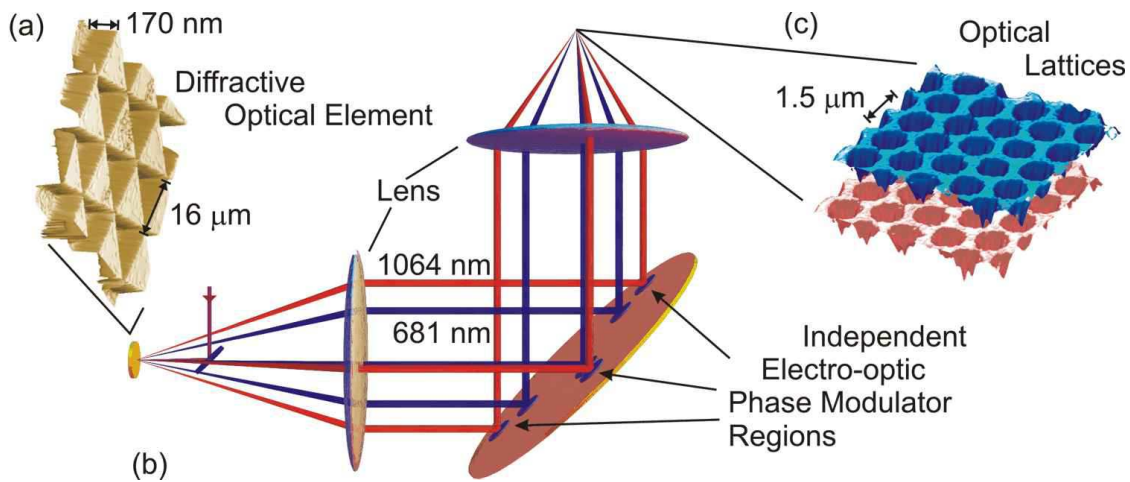}
\end{center}
\caption{Apparatus for generating a two-color optical lattice.
Copropagating beams at both wavelengths are incident on a
diffractive optical element (DOE) shown in (a), formed by a
photolithographed gold-coated fused silica surface consisting of a
regular array of raised equilateral triangles.  The image shown was
obtained with an atomic force microscope.  In (b), reflected light
is diffracted, primarily into three first order beams at each
wavelength in a triangular pattern.  These beams are then routed by
a pair of lenses to form a pair of triangular optical lattices shown
in (c), on the image plane of the DOE.  The relative position of the
two lattices is controlled with a set of electro-optic phase
modulators, formed by patterned deposition of mirror/electrodes onto
the rear surface of a single lithium-niobate crystal.  The lattice
structure shown was imaged with a microscope objective onto a CCD
camera.} \label{lattice_optics}
\end{figure}

In order to perform many identical computations simultaneously, it is necessary to match the lattice constants of the messenger and qubit lattices to commensurate values.  The constraints imposed by necessary lattice depth to achieve sufficiently low bit migration, off-resonant scattering rate, and independent control of each atomic species, preclude the use of lasers with commensurate wavelengths and the formation of lattices by counterpropagating beam pairs.  For this reason, we have tuned the intersection angles of the beams to match the relative propagation vectors.  We have chosen to work with the most simple two-dimensional potential which is topologically stable against changes in relative phase of the constituent beams, consisting of three beams at each wavelength whose $k$-vectors projected onto the plane $\vec{k}_{\perp i}^m$, where $m$ represents the wavelength, are equal in magnitude $k_{\perp}$ and distributed evenly on a unit circle, see Fig.~\ref{lattice_optics}.  The angle of each wavevector to the normal from the plane is chosen to be $\theta_m=\sin^{-1}(2\lambda_m/3d)$, where $d$ is the common lattice constant.  Each wavelength then creates a two-dimensional intensity pattern of the form $I(x,y)\propto 6-\sum_j \cos^2 (\sqrt{3}k_{\perp} r_j/2+\phi_j)$, and $r_j=x\cos(2j\pi/3)+y\sin(2j\pi/3)$ for $j=1,2,3$ are determined by the relative wave vectors of the beams, and $\phi_j$ are determined by the relative optical phase of the beams.  Finally, in the third direction, a single standing wave is applied by the intersection of two beams at a small angle, producing a lattice constant similar to that in the plane.

\begin{figure}[htpb]
\begin{center}
\includegraphics [width=0.45\columnwidth,keepaspectratio]{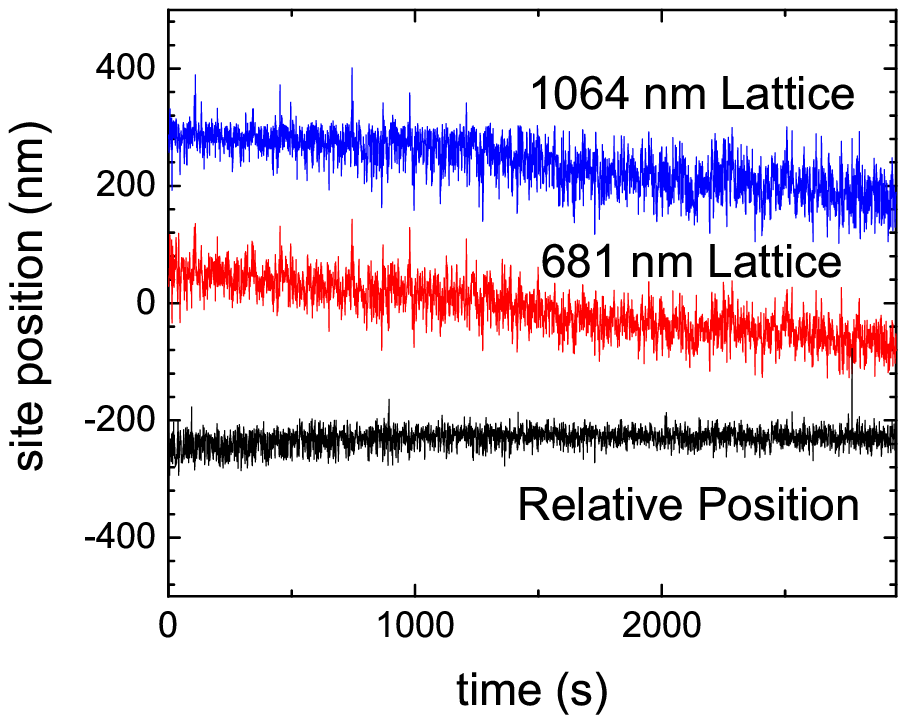}
\includegraphics [width=0.45\columnwidth,keepaspectratio]{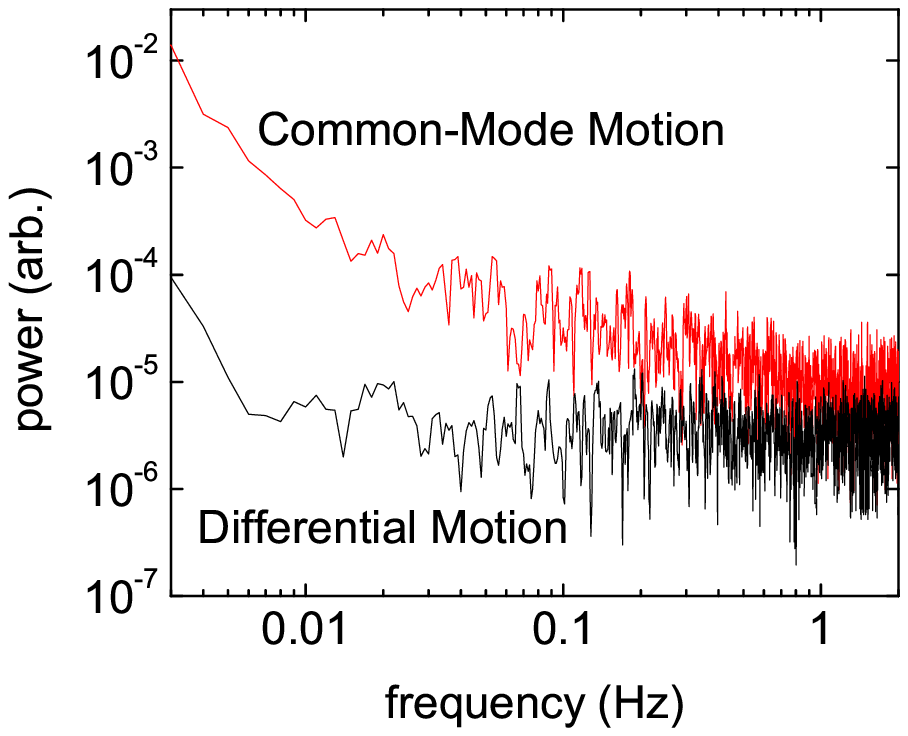}
\end{center}
\label{lattice_perf}\caption{Translational stability of the two-color optical lattice potential. The intensity distributions for both color lattices were recorded simultaneously by imaging onto a CCD camera with a microscope objective. In figure (a), a time series of motion for each color optical lattice is presented, demonstrating a single-color root-mean-square displacement of 92~nm over a 3000~s measurement time.  The relative motion is substantially smaller due to the use of common-mode optics, with a RMS displacement of 26~nm over the same time period.  A typical experiment cycle is expected to require on the order of 10~s.  In part (b), a power spectrum is presented, demonstrating cancelation of motional noise over a wide range of frequencies.}
\end{figure}

Precise and stable tuning of intersection angles and relative beam phases $\phi_j$ can be achieved by the use of a combination of diffractive and refractive optics.  In this scheme, a two-dimensional diffraction grating can be employed as a three-way beam splitter whose output beam angles are dependent on wavelength, with precisely the relation necessary to generate matched lattice constants at arbitrary input wavelengths.  These diffracted beams can be mapped onto the location of the atoms using refractive imaging techniques, as shown in Fig.~\ref{lattice_optics}; we note that only three selected spatial frequencies are allowed to propagate through the imaging lenses.  By employing only ``common-mode'' optics, through which all beams at each wavelength pass, a highly phase-stable optical setup can be constructed, largely insensitive to mount vibration and drift.  A time-series of the minimum location for a two-color lattice is shown in Fig.~\ref{lattice_perf} as recorded by imaging the lattice onto a CCD with a microscope objective, demonstrating a stability of 92~nm over 3000~s.  The differential translational stability is measured to be 26~nm over 3000~s.  This is to be compared with the site spacing $d=1.5\,\mu$m and anticipated oscillator length of the cesium atom in the lattice of 82~nm, and for lithium 165~nm.

To control the relative position of cesium and lithium atoms in the lattices, we insert optical phase modulators to control the relative phases of the beams for at least one wavelength; see Fig.~\ref{lattice_optics}.  For this purpose, we have chosen electro-optic phase modulators for their high bandwidth and relative precision.  In order to retain as much as possible a common-mode optical setup, we integrate several longitudinal electro-optic modulators into a single, large diameter lithium-niobate crystal wafer by patterning multiple electrodes onto its surfaces; see Fig.~\ref{lattice_optics}.  To lower the electrical potential difference necessary to effect a change in optical path length, the modulators are used in double-pass, with the rear electrodes serving also as mirror coatings, realized by deposition of a patterned layer of silver onto the back surface of the lithium-niobate crystal.  The front electrodes consist of a single indium-tin-oxide coating held at a common potential.

A major concern in preserving the coherence of atomic internal states is to provide a potential which is independent of the internal state of the atom.  However, for detunings not large compared to the fine structure splitting, electron spin is not decoupled from the effect of the driving optical field, and one must account for the internal state-dependent light-shift introduced by polarization gradients.  In the context of this experiment, this has two potentially important consequences.  First, it leads to a potential dephasing mechanism in the presence of inhomogeneous light fields.  In the limit of low magnetic field, the internal states of free atoms are eigenstates $|F,m_F\rangle$ of the total angular momentum $F$, and the vector light shift takes the form of an effective magnetic field $\mathbf{B}_{eff}$ \cite{deutsch1998}, proportional to the constant $D_{FS}=(\Delta_{3/2}-\Delta_{1/2})/(\Delta_{3/2}/2 + \Delta_{1/2})$, determined by the detunings $\Delta_{3/2},\Delta_{1/2}$ of the lattice light from the two excited state fine structure components; for lithium and lattice light at $\lambda=681nm$, $D^{681nm}_{FS}=1.4 \times 10^{-4}$, however for cesium $D^{681nm}_{FS}=-0.11$ and $D^{1064nm}_{FS}=0.19$.  We note that a lattice formed by beams with parallel polarizations will exhibit no such state-dependent light shift.  While this is possible in a planar geometry with intersection angles $\theta_m=\pi /2$, smaller intersections lead to a nonzero effective magnetic field. For optimally chosen linear polarizations, $\mathbf{B}_{eff}$ varies in space, exhibiting zero magnitude but nonzero gradient at the location of the scalar potential minimum.  Assuming application of a substantially larger external magnetic field in the $z$-direction, only the gradient of the component in this direction is relevant.  This is on order of $1$ kG/cm for cesium due to each lattice, which leads to a spatial separation of the minima for different internal states orthogonal to the direction of optical polarization, but introduces no shift in ground state energy to lowest order.  The peak value of $|\mathbf{B}_{eff}|$ reaches a maximum at the scalar potential minimum of order 200~mG for cesium and 100$\,\mu$G for lithium.

The amount of decoherence resulting from the polarization gradients above depends critically on the chosen internal states, as well as the magnitude of applied magnetic field.  At low fields, it is possible to place cesium atoms only in superpositions of the clock states $|F=3,\,m_F=0\rangle$ and $|F=4,\,m_F=0\rangle$, and lithium atoms in states of equal magnetic moment, in which case we expect to be largely insensitive to deleterious magnetic field and polarization gradient inhomogeneities. It is important to note that the cesium clock states remain good quantum states to relatively high magnetic fields of order 100~G, whereas the chosen lithium states enter the high-field regime at smaller fields of order 10~G.  However, the relatively small fine-structure splitting of lithium assures a modest influence of polarization gradients in all cases.

%% file: conclusion.tex
\section{Conclusion}
\label{conclusion}

We have presented a scheme for scalable quantum information processing based on two-species of ultracold atoms held in controlled bichromatic optical lattice potentials, including methods to entangle $^6$Li and $^{133}$Cs atoms locally through coupling to bound $^6$Li-$^{133}$Cs molecules, and methods to transport entanglement to distant atoms through multiple quantum manipulations. We have identified simple quantum logic gate operations possible in this scenario. Methods are based on the production of translatable optical lattices at two wavelengths with identical structure, for which we have demonstrated a novel realization utilizing diffractive optics and electro-optic modulation.  We have discussed gate operations in detail, identifying necessary timescales for entangling via a molecular state and transporting atoms adiabatically.  This compares favorably to the expected coherence time, including the effects of off-resonant scattering, qubit tunneling, external field instabilities and state-dependent light shifts. Finally, we have analyzed the effects of realistic experimental uncertainties to ascertain expected fidelities, and compared this to measured errors in lattice construction; with incremental improvement in passive stability, fidelities of $>$97\% might be achievable in entangling nearby qubits. 